\documentclass[
	twocolumn,
	amsmath,
	amssymb,
	prl,
	aps,
	nofootinbib,
	]{revtex4-2}
\usepackage{graphicx}
\usepackage{xcolor}
\usepackage{amsmath}
\usepackage{amssymb}
\usepackage{booktabs}
\usepackage{titlesec}

\begin{document}

\title{Transients and multiperiodic responses: 
a hierarchy of material bits}

\author{C.M.~Meulblok}
\author{M.~van Hecke}
\affiliation{Huygens-Kamerlingh Onnes Lab, Universiteit Leiden, P.O.~Box~9504, NL-2300 RA Leiden, Netherlands}
\affiliation{AMOLF, Science Park 104, 1098 XG Amsterdam, Netherlands}

\date{\today}

\begin{abstract}
When cyclically driven, certain disordered materials exhibit transient and multiperiodic responses that are difficult to
reproduce in synthetic materials.
Here, we show that elementary multiperiodic elements with period $T = 2$ --- togglerons --- can serve as building blocks for such responses. We experimentally realize metamaterials composed of togglerons with tunable transients and periodic responses — including odd periods.
Our approach suggests a hierarchy of increasingly complex elements in frustrated media, and opens a new strategy for rational design of sequential metamaterials.
\end{abstract}

\maketitle
The sequential response of frustrated materials to cyclic driving characterizes their memory effects \cite{SlotterbackPRE2012,PaulsenPRL2014,AdhikariEPJ2018,KeimRMP2019,PaulsenPRSA2019,YehPRL2020,KeimPRR2020,BensePNAS2021,ShohatPNAS2022,PaulsenAR2025,RegevPRE2021}
and provides a window on the organization of their underlying energy landscape \cite{RegevPRE2021}. Materializing such complex sequential responses in designer materials allows sequential shape morphing \cite{SilverbergSci2014,OverveldePNAS2015,GelebartNat2017,CelliSM2018,CoulaisNat2018,MelanconAdvFunM2022,MeeussenNat2023}
and {\em in-materia} computing \cite{SerraGarciaPRE2019,LiuPNAS2024,YasudaNat2021,KasparNat2021,ChenNat2021,ElHelouNat2022,MeulblokArXiv2025,KwakernaakPRL2023}. 
Transient and multiperiodic responses exemplify exotic sequential behavior: although multistable systems with finitely many states must eventually reach a periodic orbit, transients \( \tau > 0 \) or periods \( T > 1 \) are rare
\cite{BensePNAS2021,KwakernaakPRL2023,SchreckPRE2013,RegevPRE2013,RoyerPNAS2015,KawasakiPRE2016,LavrentovichPRE2017,NagasawaSM2019,KeimSciAdv2021,LindemanSciAdv2021,SzulcJCP2022}. Hence, understanding how these phenomena can be modeled, modified and rationally designed would provide insight into the complex physics of frustrated materials and may open routes to smart materials with advanced sequential functionalities \cite{LiuPNAS2024,YasudaNat2021,KasparNat2021,MelanconAdvFunM2022,kampArXiv2024}. 

Frustrated materials can often be modeled as collections of interacting multi-state elements
\cite{RegevPRE2021,BensePNAS2021,ShohatPNAS2022,KeimSciAdv2021,LindemanSciAdv2021,SzulcJCP2022,MunganPRL2019,TerziPRE2020,RegevPRE2021,HeckePRE2021,TeunisseArXiv2024,ElElmiSM2024}.
Binary spins and hysterons
are popular choices for these elements. For spins, their phase depends solely on the instantaneous driving parameter $\varepsilon$, whereas for hysterons, it additionally depends on the driving history (Fig.~1a-b).
Interactions between elements are crucial for
nontrival transients and multiperiodic behavior;
for example, $T=2$ orbits require six interacting spins or three interacting hysterons \cite{KeimSciAdv2021,DeutschPRL2003} (see SI). 

Here we 
introduce {\em togglerons}, elements that return to their initial state after two driving cycles, as effective building blocks to understand and realize both transient and multi-periodic behavior(Fig.~1c).
Spins, hysterons and togglerons are part of a hierarchy, where elements of higher rank mimic multiple interacting elements of a lower rank; for example, 
two interacting spins can form a single hysteron, and togglerons can be modeled by groups of spins of hysterons  (Fig.~1d; see SI).
This motivates exploring the tradeoff between the complexity of individual elements and the number of elements required to achieve specific behaviors and orbits.

\begin{figure}[tb]
    \centering
    \includegraphics[width=\linewidth]{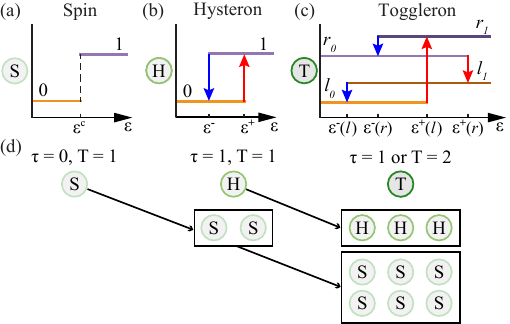}
    \caption{Material Bits.
    (a-c) Transitions and switching thresholds for spins, hysterons and togglerons. Even though togglerons have more than two states, we refer to these as material bits.
    (d) Hierarchy of material bits: Two interacting spins (S) can model a hysteron (H); six interacting spins or three interacting hysterons can model a toggleron (T).}
    \label{fig:01}
\end{figure}

We focus on mechanical togglerons and  materialize these with cyclically compressed beams that buckle in a constrained environment. We investigate interacting togglerons and show how to rationally design their properties and interactions for
targeted orbits. We finally materialize a toggleron-metamaterial which, depending on the amplitude of the cyclic drive, exhibits orbits with $(\tau,T)=(2,2),(0,4)$ or $(1,3)$.
Our hierarchical framework, where clusters of simple interacting elements form aggregate units with complex sequential responses, enables the design of smart materials for sensing, soft robotics, and sequential {\em in materia} computing.

\textit{Togglerons.---}
Conceptually, a toggleron requires four distinct configurations that we label \( l_0, l_1, r_0\) and \(r_1 \), effectively doubling the usual '0' and '1' states of a hysteron into 'left' and 'right' leaning flavors (Fig.~1c).
Over the course of two driving cycles, such toggleron follows a specific transition sequence: During the first up-sweep of \( \varepsilon \), it switches from configuration \( l_0 \) to \( r_1 \) at 
switching threshold
\( \varepsilon^+(l) \). On the subsequent down-sweep, it transitions from \( r_1 \) to \( r_0 \) at \( \varepsilon^-(r) \). In the next cycle, the configuration evolves from \( r_0 \rightarrow l_1 \rightarrow l_0 \), with switching events at \( \varepsilon^+(r) \) and \( \varepsilon^-(l) \), respectively (Fig.~1c). By ensuring that both down switching thresholds 
$ \varepsilon^-$ are smaller than both up switching thresholds $\varepsilon^+$, togglerons feature a hysteretic regime where all four configurations are stable and exhibit a robust $T=2$ orbit under cyclic driving \cite{TenWoldeMMT2024}.

We materialize mechanical togglerons using slender beams (\( L_0 =100\) mm) constrained by lateral boundaries with two protrusions, or tips \cite{MeulblokEML2025}. Under cyclic compression of the beam, these tips induce snapping between its left- and right-buckled configurations. 
We non-dimensionalize all lengths by \( L_0 \), including the width \( t \) and center position \( x_b \) of the beam. 
The lateral boundaries are fixed to the bottom plate, and are parameterized by their heights $z_b$ and $z_t$ and horizontal position $x_l$ and $x_r$ (Fig.~\ref{fig:02}a; see Appendix A). 
The system is driven by quasistatic compression cycles, with strain \( \varepsilon \) varying between \( \varepsilon_m \) and \( \varepsilon_M \).
Crucially, we take \( \varepsilon_m > \varepsilon_b \), where \( \varepsilon_b \) is the buckling strain, which ensures that the beam remains buckled and history dependent \cite{KwakernaakPRL2023}.

To demonstrate the targeted sequential evolution, a toggleron is initialized in the left buckled configuration \( l_0 \), and the mid-point deflection \( x_b \) is tracked under cyclic driving. Our data show that the beam undergoes four distinct irreversible transitions over two cycles—one during each up or down sweep of \( \varepsilon \)—and returns to its initial configuration at the end of the second cycle, as required (Fig.~\ref{fig:02}b, Movie 1).

We now discuss these snapping transitions. 
The transition between $l$ and $r$ configurations during increasing $\varepsilon$ is triggered by the buckled beam coming into 
contact with the tips
(Fig.~\ref{fig:02}bi). This initially leads to a smooth deformation of the beam into a higher order mode (Fig.~\ref{fig:02}bii), and eventually to irreversible
snapping of the beam to a right-leaning configuration $r_1$ (Fig.~\ref{fig:02}biii). 
Under decreasing $\varepsilon$,  the beam initially deforms smoothly (Fig.~\ref{fig:02}biv) before a (subtle) unsnapping transition (Fig.~\ref{fig:02}bv).
Despite its small displacement (Fig.~\ref{fig:02}c, inset), this transition is irreversible and crucial---under compression, configurations \( r_1 \) and \( r_0 \) (or \( l1 \) and \( l_0 \))  evolve qualitatively differently (Fig.~1c, Fig.~\ref{fig:02}c, see SI).

The rationale for dual-tip boundaries is as follows.
To allow lateral motion of the beam,
the tips of the left and right boundaries should be placed to the left (\( x_l < 0 \)) and right (\( x_r > 0 \)) of the beam. 
Yet, these boundaries must {induce snapping} of the beam. Specifically, if a beam is initially buckled to the left, increasing compression should trigger a snap-through transition to the right.
For a single tip boundary, such snapping always requires
$x_l>0$ \cite{Pandey14,Vangbo98,Chen21,MeulblokEML2025}. However,  the dual-tip geometry allows to
exert a moment on the beam, and when properly designed, induces snapping from $l$ to $r$ for $x_l<0$, thus powering the toggling motion of the beam \cite{MeulblokEML2025}.

The locations of the tips control the critical thresholds; in the example above, the toggleron is left-right symmetric ($x_l\!=\!-x_r$) so that  
$\varepsilon^+(l)=\varepsilon^+(r)$ and
$\varepsilon^-(l)=\varepsilon^-(r)$. By breaking left-right symmetry, and varying the vertical locations of the pusher tips ($z_b$ and $z_t$) all switching thresholds can be tuned independently \cite{MeulblokEML2025}.

\begin{figure}[t]
    \centering
    \includegraphics[width=\linewidth]{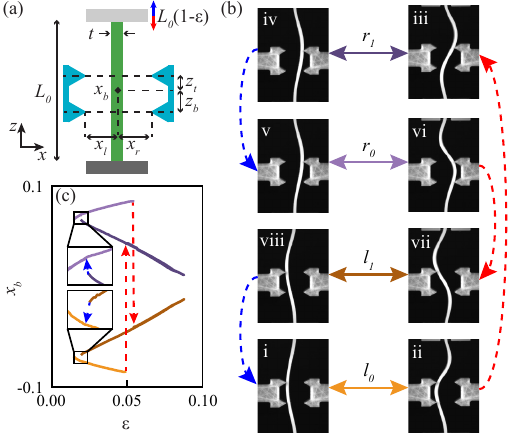}
    \caption{Geometry and phenomenology of togglerons. 
    (a) We realize togglerons using a slender beam (green) constrained by fixed lateral boundaries (blue). The black circle at the center of the beam at rest indicates the origin of the \( (x,z) \)  coordinate frame.   
    (b) Evolution of a toggleron during two compression cycles ($(\varepsilon_m,\varepsilon_M)= (0.015,0.088$),  $( t, z_t, z_b, x_l,x_r)=( 0.025,0.02,-0.16,-0.08,0.08)$. 
    The snapshots are taken just before and after the irreversible transitions during compression (red) and decompression (blue).
Colored bidirectional arrows indicate reversible beam evolution, corresponding to the four configurations $r_1,r_0,l_1$ and $l_0$. 
    (c) Corresponding  mid-point deflection $x_b$. Transitions occur rapidly, i.e., at fixed $\varepsilon$ (vertical dashed arrows). The insets illustrate the irreversible transition at $\varepsilon^-(0)$.
    }\label{fig:02}
\end{figure}

\begin{figure*}[t]
\centering
\includegraphics[]{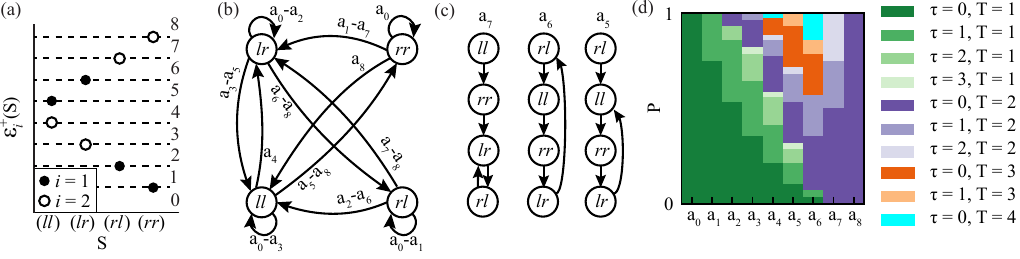}
\caption{
Evolution of two interacting togglerons.
(a) Example of the ordering of the eight switching thresholds $\varepsilon^+_i(S)$ (symbols) which define nine strain intervals (0-8).
(b) Corresponding FSM diagram. 
(c) Three example orbits under cyclic driving with $a_7$, $a_6$, and $a_5$.    (d) Normalized stacked bar chart for $(\tau,T)$. 
    }\label{fig:theory}
\end{figure*}

\textit{Elementary transients. ---} 
Togglerons exhibit both 
multiperiodic and transient
behavior.
Consider periodic driving of (asymmetric) togglerons for
 $\varepsilon_m<\varepsilon^-$ so that  the $l_1\!\rightarrow \!l_0$ and $r_1\!\rightarrow \!r_0$ transitions always occur. The (relative) values of  $\varepsilon^+(l), \varepsilon^+(r)$ and $\varepsilon_M$ then control the evolution after each full driving cycle, allowing
to drop the '0' and '1' indices and to characterize the toggleron as $l$ or $r$.
 When $\varepsilon_M\!<\!\varepsilon^+(l)$
and $\varepsilon_M\!<\!\varepsilon^+(r)$, the toggleron remains stuck and $(\tau,T) = (0,1)$. When
$\varepsilon_M$ exceeds both switching thresholds, $(\tau,T) = (0,2)$. For $\varepsilon^+(l)<\varepsilon_M<\varepsilon^+(r)$, which can be realized by taking $|x_l|<|x_r|$, the evolution depends on the initial state: Starting from $r$, the toggleron  remains stuck, while starting from $l$, it first evolves to $r$ and then remains stuck, leading to an elementary transient: $(\tau,T) = (1,1)$.

\textit{Interactions and complex orbits. ---} 
To explore a broader range of transients and multi-periodic orbits\cite{KwakernaakPRL2023,SchreckPRE2013,RegevPRE2013,RoyerPNAS2015,KawasakiPRE2016,LavrentovichPRE2017,NagasawaSM2019,KeimSciAdv2021,LindemanSciAdv2021,SzulcJCP2022}, we consider two interacting togglerons. We denote their collective states as
$S=(ll),(lr),(rl)$ or $(rr)$, and note that these
four states constrain the possible transients and periods to \( \tau + T \leq 4 \).
We encode interactions
in eight the state-dependent switching thresholds $\varepsilon^+_i(S)$, where $i=1,2$ labels the toggleron.

Specific orbits depend on the interval 
in which
$\varepsilon_M$ is chosen (Fig.~\ref{fig:theory}a)
\cite{TeunisseArXiv2024,LiuPNAS2024}. First, the thresholds divide $\varepsilon$ in nine intervals and
choose one representative value 
$\varepsilon_M^k$ in each interval that we label $k=0,1,\dots,8$ (Fig.~\ref{fig:theory}a) 
\cite{LiuPNAS2024}. 
We then define nine input characters \( a_k\) representing full driving cycles 
\( \varepsilon: \varepsilon_m \uparrow \varepsilon^k_M \downarrow \varepsilon_m \). Together, this allows to determine the evolution of any initial state $S_0$ after one driving cycle, $S_1=a_k(S_0)$, and to encode the evolution in a finite state machine (FSM) diagram (Fig.~\ref{fig:theory}b) \cite{LiuPNAS2024}.  

The FSM enables rapid classification of transients and multi-periodic responses under cyclic driving.
For example, for $\varepsilon_M=\varepsilon^7_M$,
the response can be tracked by following the transitions in the FSM under repeated compression cycles $a_7$. Starting from $S_0=(ll)$, the system exhibits a transient of two before entering an orbit with $T=2$: $(ll)\xrightarrow{a_7}(rr)\xrightarrow{a_7}(lr)\xrightarrow{a_7}(rl)\xrightarrow{a_7}(lr)\xrightarrow{a_7}\dots$ (Fig.~\ref{fig:theory}b,c). Other initial states yield $(\tau=1,T=2)$ and $(\tau=0,T=2)$. Similarly, for $\varepsilon_M=\varepsilon^6_M$, a cycle of maximal length four
$(\tau=0,T=4)$ is observed. Strikingly, for $\varepsilon_M=\varepsilon_M^7$, a cycle of period {\em three} emerges (Fig.~\ref{fig:theory}c).

Since the possible transients and periods only depend on the ordering of these thresholds,
we determine $(\tau,T)$ for 
all $8!$ orderings and explore their statistics
(Fig.~\ref{fig:theory}d). 
The combination of non-symmetric elements ($\varepsilon^+(l) \neq \varepsilon^+(r)$) and interactions enables the realization of all allowed combinations of $\tau$ and $T$, including multiperiodic orbits with odd periods.
Together, our data and examples illustrate that interacting togglerons exhibit a wide range of transients and multiperiodic orbits which depend on driving amplitude and initial state. 

{\em Rational design of orbits.---} For each orbit, conditions on the ordering of the switching thresholds can straightforwardly be designed. We illustrate this for 
the orbit $(rl)\xrightarrow{a_k}(ll)\xrightarrow{a_k}(rr)\xrightarrow{a_k}(lr)\xrightarrow{a_k}(ll)\xrightarrow{a_k}\dots$. Each transition, e.g., $(rl)\xrightarrow{a_k}(ll)$ produces two design inequalities,
i.e., $\varepsilon^+_1(rl)<\varepsilon_M^k$ and $\varepsilon^+_2(rl)>\varepsilon_M^k$. 
These transitions collectively impose up to eight inequalities and define one or more admissible orderings of the switching thresholds and associated ranges for $\varepsilon_M$.
Note that for the specific design problem of finding switching thresholds that exhibit the three orbits shown in Fig.~\ref{fig:theory}c, we only require 
the specific ordering of the four largest thresholds, i.e., a partial order of the thresholds \cite{TeunisseArXiv2024}.
Hence, specific (combinations of) orbits, transients and periods can be designed by controlling the switching thresholds. 

This simple design strategy highlights the utility of togglerons. Realizing similar orbits with spins and hysterons require solving a larger, more intricate set of design inequalities~\cite{HeckePRE2021,KeimSciAdv2021,LiuPNAS2024,TeunisseArXiv2024,LindemanSciAdv2025}. This is primarily due to their limited response in isolation - hysterons can exhibit at most a $\tau=1$ response, while spins lack any transient behavior. 
By employing a more versatile unit such as a toggleron, which can exhibit both multiperiodic and transient responses, the design of more complex orbits becomes markedly more straightforward.

\begin{figure*}[t]
    \centering
    \includegraphics[]{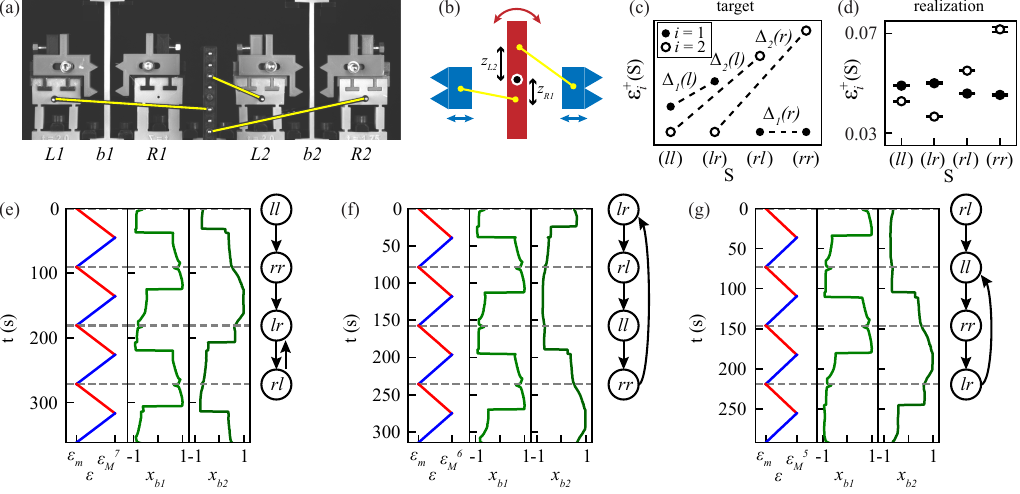}
    \caption{Metamaterial consisting of
two interacting togglerons.
    (a) Geometry highlighting the connections from the flexible boundaries $L1,L2$ and $R2$ to a central rotating arm (yellow; see SI).
    (b) Schematic representation of the  coupling between boundaries $L2$ and $R1$.
    (c) Target ordering of the switching thresholds. 
    (d) Realized switching thresholds for our specific design (for our specific design, see SI).
    (e-g) Evolution of the mid-point deflections $x_{b1}$ and $x_{b2}$ during four compression cycles with amplitude $\varepsilon_M^7=0.056$, $\varepsilon_M^6=0.050$, and $\varepsilon_M^5=0.047$, showing orbits with $\{\tau,T\}$=$\{2,2\}$, $\{0,4\}$ and $\{1,3\}$. 
    }\label{fig:twoTexp}
\end{figure*}

{\em Materializing targeted orbits.---}
To demonstrate the feasibility of this approach, we create a metamaterial that realizes
the three orbits shown in Fig.~\ref{fig:theory}c.  
This metamaterial consists of
two asymmetric, interacting togglerons,  where flexible lateral boundaries mediate their interactions (Fig.~\ref{fig:twoTexp}a). 
We control the left-right symmetry by the spacing (at rest) between beam and boundaries
($L_i, R_i$) (Fig.~\ref{fig:twoTexp}a,b). 
Interactions are implemented by placing 
the lateral boundaries on a weakly deformable structure. 
Hence, when, e.g., 
beam one is in contact with its right pusher it displaces it, and by coupling the right pusher of beam one with, e.g., the left pusher of beam two, interactions between the togglerons are realized. We
couple the boundaries 
via a rotating structure, so that 
magnitude and sign of the interactions 
are controlled by the stiffness of the pusher support and the arms of the rotating structure.
(Fig.~\ref{fig:twoTexp}b).

Materializing
the target orbits  requires a partial order of the thresholds, where the four largest thresholds are 
$\varepsilon_1^+(ll)<\varepsilon_1^+(lr)<\varepsilon_2^+(rl)<
\varepsilon_2^+(rr)$
(Fig.~4c).
The differences between, e.g., 
$\varepsilon_1^+(ll)$ and
$\varepsilon_1^+(lr)$ (dashed lines in Fig.~4c)
necessitate interactions between the togglerons, which we model as 
\begin{equation}
\varepsilon_i^+(S)=e_i^+(s_i)+\Delta_i(s_i) B(s_j).~
\end{equation}
Here $B(l):=-1$, $B(r):=1$, and
$e_1^+(l)$, $e^+_1(r)$, $e^+_2(l)$ and $e^+_2(r)$
denote the `bare' thresholds of the isolated togglerons.
The interaction strengths are given by
$\Delta_i(s_i)$, where, e.g.,
$\Delta_1(l):=(\varepsilon_1^+(lr)-\varepsilon_1^+(ll))/2$ 
denotes how strong toggleron two modifies the threshold of phase $l$ of 
toggleron one. Hence, 
the targeted ordering implies three conditions:
{\em(i)} $\Delta_1(r)\approx 0$;
{\em(ii)} $\Delta_1(l)>0,
\Delta_2(l)>0$ and
$\Delta_2(r)>0$;
{\em(iii)} $\Delta_1(l) \ll \Delta_2(l)< \Delta_2(r)$  (Fig.~4c). 

We implement these conditions by selecting appropriate design parameters. First, condition {\em(i)} is satisfied by boundary $R_1$ being
uncoupled from other boundaries.
Second, the sign of the interactions (condition {\em(ii)}) implies that the coupling makes 
boundaries $L_1$, $L_2$ and $R_1$ all move in and outward together, which requires that $z_{L1}/z_{R2}>0$ and $z_{L1}/z_{L2}<0$. 
Third, the non-symmetric nature of the interactions (condition {\em(iii)})
is realized by choosing beam one thicker than beam two, and setting $z_{L1}<z_{L2}<z_{R2}$.  
Guided by these considerations, and iterative adjustments of $x_{Li}$ and $x_{Ri}$ to finetune the switching thresholds, we select metamaterial parameters that meet the target ordering (Fig.~\ref{fig:twoTexp}d; see SI).

To demonstrate that this 
metamaterial
exhibits the three target orbits, it is initialized in the appropriate initial state and its evolution is followed during cyclic sweeps of $\varepsilon$ between
$\varepsilon_m=0.0125$ and $\varepsilon_M^5$, $\varepsilon_M^6$ and
$\varepsilon_M^7$ (Fig.~4efg).
First, for the largest driving ($k=7$), the metamaterial
exhibits a transient of two and then locks in period-two orbit where both togglerons toggle each driving cycle $(lr)\rightarrow(rl)\rightarrow (lr) \rightarrow ...$ (Fig.~\ref{fig:twoTexp}c; see Movie 2). 
Second, for intermediate driving ($k=6)$, the metamaterial exhibits the longest possible orbit of period $T=4$
(Fig.~\ref{fig:twoTexp}d; see Movie 3). Finally, pushing the limit of our experiment, when $\varepsilon_M$ is chosen in the narrow interval corresponding to $k=5$, the metamaterial exhibits a short transient and then a period three response
(Fig.~\ref{fig:twoTexp}e; see Movie 4).
Hence, 
a {\em single} metamaterial exhibits
all three target orbits.

\textit{Discussion and Outlook.---} 
We introduced togglerons, elements which
repeatedly evolve over two driving cycles,
as a natural step in a hierarchy of elements that starts with binary spins and hysterons.
Their intrinsic multiperiodic nature facilitates complex orbits with long transients or large periods - something which requires much larger groups of interacting spins or hysterons  \cite{PaulsenArXiv2024,KeimSciAdv2021} - thus opening up routes to sequential mechanisms, soft robotics and mechanical information processing \cite{kampArXiv2024,KwakernaakPRL2023,LiuPNAS2024,YasudaNat2021,KasparNat2021}.

In the context of amorphous systems, our work suggests that 
large periods and long transients may arise from interactions of local clusters of particles which in isolation exhibit only short periods or transients \cite{SchreckPRE2013,RegevPRE2013,RoyerPNAS2015,KawasakiPRE2016,LavrentovichPRE2017,NagasawaSM2019,KeimSciAdv2021,LindemanSciAdv2021,SzulcJCP2022}. 
We suggest that composite building blocks such as the togglerons,
or the pairs of frustrated hysterons 
that recently have been shown to
capture non-trivial
'latching' memories \cite{LindemanSciAdv2025} are not the end of the hierarchy. In particular for
hierarchically organized materials,
including glasses,
exciting questions arise: which composite elements are formed, which of these are
statistically dominant, and how far does this hierarchy extend?

\textit{Acknowledgements} \begin{acknowledgments}
{We thank M. Teunisse for 
analyzing spin and hysteron systems, and L. Kwakernaak, H. Bense, and T. Brandt for fruitful discussions and J. Mesman for technical support.}
\end{acknowledgments}

\bibliographystyle{apsrev}

\cleardoublepage

\setcounter{equation}{0}
\setcounter{figure}{0}
\setcounter{table}{0}
\setcounter{page}{1}
\makeatletter
\renewcommand{\theequation}{S\arabic{equation}}
\renewcommand{\thefigure}{S\arabic{figure}}
\renewcommand{\thetable}{S\Roman{table}}

\widetext
\begin{center}
\textbf{\large Supplemental Information}
\end{center}

\section{Hierarchy of material bits}

Here, we detail how higher-rank material bits can be constructed by composing  interacting 
material bits of lower rank.
We consider three cases: {\em (i)} Interacting binary spins that form a hysteron; {\em (ii)} Interacting hysterons that form a toggleron; {\em(iii)} Interacting spins that form a toggleron.
We consider pairwise additive interactions~\cite{HeckePRE2021,TeunisseArXiv2024,KeimSciAdv2021,LindemanSciAdv2021,LindemanSciAdv2025}:
\begin{align}\label{cij}
    U^\pm_i(S)=u_{i}^\pm - \sum_{j\ne i} c_{ij}s_j,
\end{align}
where $U^\pm_i(S)$ denotes the switching thresholds of element $i$ in the collective state $S=(s_1s_2\dots s_N)$. For the case of binary spins,
$u_i^+=u_i^-$ and $U_i^+(S)=U_i^-(S)=$. 
The state-dependent switching thresholds are composed of the bare switching fields 
$u_{i}^\pm$ and the interaction term $- \sum_j c_{ij}s_j$, and we consider both symmetric
($c_{ij}=c_{ji}$) and asymmetric interactions ($c_{ij}\neq c_{ji}$).

We first show that two spins with symmetric or asymmetric interactions can form a hysteresis loop, and that this requires avalanches where multiple spins switch simultaneously. While the argument is conceptually simple, we are not aware of any prior work in which it has been explicitly presented.
Hence, hysterons can be composed of two interacting spins. We then consider interacting hysterons that exhibit a multiperiodic ($T=2$) loop, thus forming a toggleron. From numerical sampling it was know that this requires
at least three hysterons, and that for symmetric interactions, four hysterons are needed \cite{HeckePRE2021,KeimSciAdv2021,PaulsenArXiv2024}. Here we present a systematic exploration using recently developed techniques that does not rely on sampling and is able to find all possible orbits
\cite{TeunisseArXiv2024,BaconnierArXiv2025}.
We finally consider how interacting spins can form multiperiodic orbits, and explicitly construct a 
$T=2$ orbit - toggleron - using six interacting spins.

\subsection{(i) Interacting spins forming a hysteron}
We now explore how binary spins can exhibit hysteretic behavior similar to a hysteron. First, we note that transitions in which one spin flips can not exhibit hysteresis (this follows directly from Eq.~(\ref{cij})). 
However, when avalanches are taken into account, 
pairs of interacting spins can produce hysteretic transitions, thus effectively forming a hysteron.

For symmetric interactions, a pair of interacting spins is characterized by three parameters: the bare switching fields $u_1$ and $u_2$, and the interaction strength $c_{12}=c_{21}=c$. Without loss of generality, we assume $u_1 > u_2$, and consider four parameter regimes given by 
the pair of inequalities: 
\begin{align}
    u_1 - u_2 &\gtrless |c|, \\
    c &\gtrless 0.
\end{align}
We have verified that the only possible avalanches for symmetric interactions 
are between
the states $S=(00)$ and $(11)$. These 
occur when $u_1 - u_2 < |c|$ and $c>0$ \cite{HeckePRE2021,KeimSciAdv2021,LindemanSciAdv2021,TeunisseArXiv2024}. 
Then, by choosing
$c>u_1-u_2$, the spin-pair switches from state 
$S_1=(00)$ to $S_2=(11)$ when $U$ exceeds
$U^+(00)= u_2$ or $(11)\rightarrow(00)$
when $U$ falls below $U^-(11) = u_1 - c$. Hence, two spins form a hysteron with states $S_1$ and $S_2$, and bare hysteron switching fields $u_h^+ = u_2$ and $u_h^- = u_1-c$.
For asymmetric interactions ($c_{12}\neq c_{21}$), a spin pair can also exhibit hysteretic avalanches between $S=(01)$ and $(10)$, when $c_{ij}<0$ and $|c_{21}|>u_1-c_{12} - u_2$, and where, the spin-pair forms a hysteron with effective switching fields:
$u^+=u_1-c_{12}$ and $u^-=u_1$.
Hence, there are at least two routes for pair of spins to form a hysteron.

\subsection{(ii) Multiperiodic behavior with interacting hysterons}
Interacting hysterons have been studied intensely in recent years using numerical sampling techniques, providing examples of $T=2$ multiperiodic responses that thus form togglerons \cite{HeckePRE2021,KeimSciAdv2021,LindemanSciAdv2021}.
Here, we systematically explore multiperiodic orbits adopting methods from \cite{TeunisseArXiv2024,BaconnierArXiv2025}. We first construct, for $N$ hysterons, all fundamental multiperiodic orbits, and then systematically check which of these are realizable.
Fundamental multiperiodic orbits
are repeating sequences of states connected by single hysteron flips containing multiple episodes of up or down transitions, 
specified by the starting state $S^0=(s_1^0s_2^0\dots s_N^0)$ and the sequence of element flips $(\kappa^1,\kappa^2,\kappa^3,\dots)$ \cite{BaconnierArXiv2025}. We deal with timeshift and relabeling
symmetries by indexing elements according to the order in which they are flipped, and only keep orbits with initially descending order \cite{HeckePRE2021,TeunisseArXiv2024,BaconnierArXiv2025}. In addition, we order the generated multiperiodic orbits by their minimum magnetization $m:=\sum_is_i$ and filter out isomorphic orbits \cite{BaconnierArXiv2025}. 
Together, this procedure results in a complete set of fundamental multiperiodic orbits \cite{BaconnierArXiv2025,TeunisseArXiv2024}.

To check whether a fundamental multiperiodic orbit is realizable, we first construct a set of linear inequalities on the design parameters that needs to be satisfied to realize the orbit using pairwise interactions (Eq.~\ref{cij})
\cite{TeunisseArXiv2024}. Next, to ensure that such an orbit is traversed for cyclic driving of $U$ between 
$U_m$ and $U_M$, we require that the appropriate (extremal) states are both reached and stable at $U_m$ or $U_M$, yielding an additional set of linear inequalities. Solvability of both sets of inequalities is easily verified, and explicit solutions can be constructed if necessary \cite{TeunisseArXiv2024}.

\begin{figure}[ht]
    \centering
    \includegraphics[]{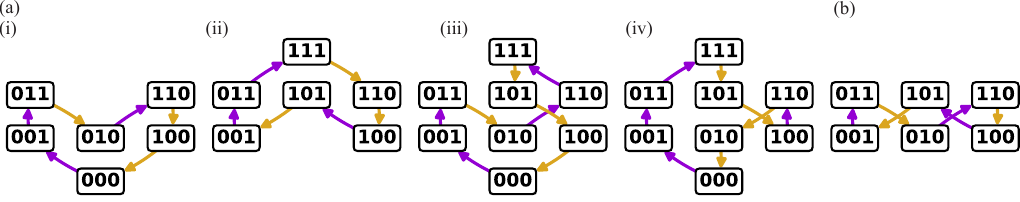}
    \caption{Realizable fundamental multiperiodic orbits with $N=3$ hysterons and asymmetric interactions ($c_{ij}\neq c_{ji}$).
    (a) $T=2$ orbits and (b) $T=3$ orbits.
    }
    \label{fig:SI01}
\end{figure}

For $N=2$ hysterons, no fundamental multiperiodic orbits exist \cite{HeckePRE2021}. For $N=3$ hysterons, asymmetric interactions can produce $T=2$ and $T=3$ orbits \cite{KeimSciAdv2021,HeckePRE2021}(Fig.~\ref{fig:SI01}, Tab.~\ref{tab:01}). 
Symmetric interactions require at least $N=4$ hysterons.

\begin{table}[htb]
\begin{tabular}{l|l|ll|l}
    &     & \multicolumn{2}{c|}{Realizable}                               &       \\
$N$ & $T$ & \multicolumn{1}{l|}{$c_{ij} = c_{ji}$} & $c_{ij} \neq c_{ji}$ & Total \\ \hline
3   & 2   & \multicolumn{1}{l|}{0}                 & 4                    & 4     \\
    & 3   & \multicolumn{1}{l|}{0}                 & 1                    & 1     \\ \hline
4   & 2   & \multicolumn{1}{l|}{62}                & 205                  & 205   \\
    & 3   & \multicolumn{1}{l|}{6}                 & 391                  & 620   \\
    & 4   & \multicolumn{1}{l|}{0}                 & 76                   & 322   \\
    & 5   & \multicolumn{1}{l|}{0}                 & 12                   & 52   
\end{tabular}
\caption{Number of realizable multiperiodic orbits for symmetric and asymmetric interactions.}
\label{tab:01}
\end{table}

\subsection{(iii) Multiperiodic behavior with interacting spins}

We now examine how binary spins can exhibit multiperiodic behavior similar to a toggleron.
Prior numerical sampling studies have found that at least five interacting spins are required to generate period $T=3$ orbits
\cite{DeutschPRL2003, KeimSciAdv2021}. Here, we focus on $T=2$ behavior, and note that we are not aware (and have not been able) to create $T=2$ orbits with five spins.

A systematic study of realizable fundamental multiperiodic orbits is not suited for spin systems, as, first, the number of orbits grows rapidly with system size, rendering exhaustive analysis impractical beyond $N>4$ \cite{BaconnierArXiv2025}, and second, fundamental orbits exclude avalanches, which are essential for irreversible transitions in spin systems \cite{TeunisseArXiv2024}. 

We construct multiperiodic orbits in spin systems by mapping spin pairs to single hysterons and using hysteron-based multi-periodic orbits. While not exhaustive, this approach offers a constructive method for generating multiperiodic behavior in systems with an even number of spins.

\begin{figure}[htb]
    \centering
    \includegraphics[]{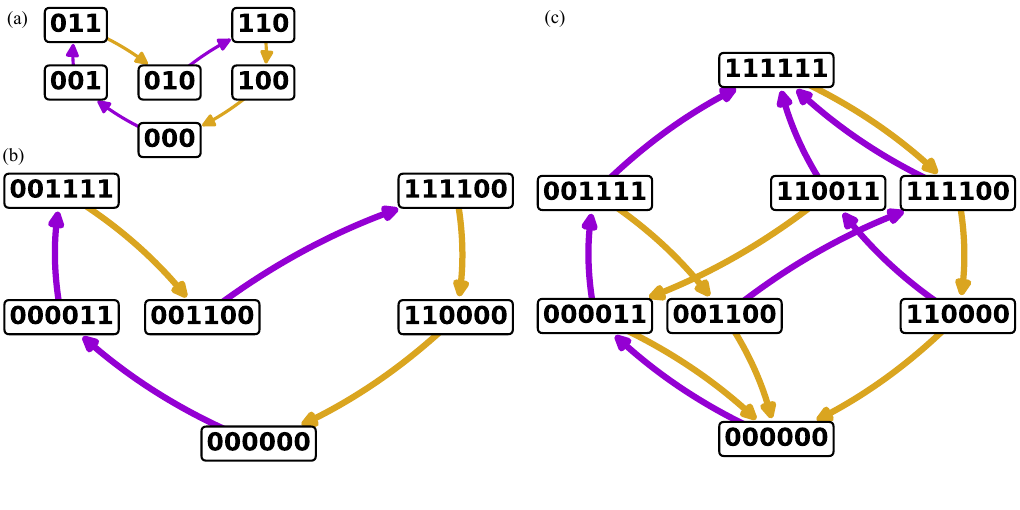}
    \caption{Construction of multiperiodic orbits in spin system. (a) Fundamental multiperiodic orbit with $T=2$, realizable with $N=3$ asymmetric interacting hysterons. (b) Resulting multiperiodic orbit with $T=2$ with $N=6$ interacting spins. (c) All transitions of the $N=6$ interacting spin system.}
    \label{fig:SI02}
\end{figure}

Our method starts 
with choosing a fundamental multiperiodic orbit (Fig.~\ref{fig:SI02}a). We generate a solution for the set of inequalities using numerical tools from \cite{TeunisseArXiv2024}. The solution has bare switching fields:
\begin{align}
    (u_1^+, u_1^-) &= ( 0.98 , -0.49 ),\\
    (u_2^+, u_2^-) &= ( 0.66 , -1.00 ),\\
    (u_3^+, u_3^-) &= ( 0.17 , -0.92 ),\\
\end{align}
and interaction matrix:
\begin{align}
    c_{ij} = 
    \begin{pmatrix}
    0 & 0.43 & -0.68 \\
    -1.00 & 0 & -0.1 \\
    -0.64 & -0.94 & 0
    \end{pmatrix}.
\end{align}
For $-1.00<U_m<-0.49$ and $0.76<U_M<1.23$ the orbit has $T=2$. 
We then map each hysteron $i$ to a symmetrically interacting spin pair, with bare switching field of spin one $u_{1,i}=u_i^--c$ and spin two  $u_{2,i}=u_i^+$.
The interactions between the spin pairs must satisfy $c_i>0$ and $|c_i|>u_i^+-u_i^-$. 
Therefore, we take $u_{1,i}=u_{2,1}+\delta$ and $c_i=u^+_i-u^+_i+\delta$, with $\delta=0.01$.
Furthermore, we choose to spread the interactions between each hysteron equally over the interacting spin pairs, such that each spin shifts the same amount and the same distance as the hysteron. This yields:
\begin{align}
    (u_{1,1}, u_{2,1}) &= (0.99, 0.98), \\
    (u_{1,2}, u_{2,2}) &= (0.67, 0.66), \\
    (u_{1,3}, u_{2,3}) &= (0.18, 0.17), \\
\end{align}
with interaction matrix (in block form):
\begin{align}
    c_{ij} = 
    \begin{pmatrix}
        0 & 1.48 & 0.215 & 0.215 & -0.34 & -0.34 \\
        1.48 & 0 & 0.215 & 0.215 & -0.34 & -0.34 \\
        -0.50 & -0.50 & 0 & 1.67 & -0.05 & -0.05 \\
        -0.50 & -0.50 & 1.67 & 0 & -0.05 & -0.05 \\
        -0.32 & -0.32 & -0.47 & -0.47 & 0 & 1.10 \\
        -0.32 & -0.32 & -0.47 & -0.47 & 1.10 & 0 \\
    \end{pmatrix}
\end{align}
Together, this produces a $T=2$ orbit (Fig.~\ref{fig:SI02}b-c). 

\section{Experimental details}
\subsection{Sample fabrication, Setup and design}

We now discuss the fabrication of the togglerons.
Experiments are performed with beams of length $L_0=100\pm0.5\,$mm, out-of-plane width of $20\pm0.5\,$mm and various in-plane thickness (see below).
Beams are fabricated by casting two-component polyvinyl siloxane elastomer (Zhermack Elite double 22 with Young's modules $0.8\,$MPa and Poisson's ratio $\approx 0.5$) into 3D-printed molds. After curing for seven days at room temperature, the beams are removed and dusted with talc powder to minimize friction and prevent sticking. 

In experiments on a single toggleron (Fig.~2 main text),
the lateral pushers are 3D printed and rigidly mounted on precision linear stages for accurate positioning and alignment.
In the experiments on pairs of togglerons (Fig.~4 main text), 
the lateral pushers are composed of three parts: 1) flexible legs, 2) pusher base, and 3) protrusions. The flexible legs are fabricated similarly to the beams, while  the pusher base and protrusions are rigid and 3D printed. The locations $x_l,x_r$ can be adjusted by sliding motion.

Cyclic driving is performed using 
a custom build compression device
that allows for accurate parallel compression of wide structures (top and bottom plates remain parallel within a slope of $<0.6~\text{mm}/\text{m}$ \cite{KwakernaakPRL2023}).
The compressive strain $\varepsilon$ is controlled by a stepper motor yielding an accuracy $\pm 0.01~\text{mm}$. We remain in the quasi-static regime during compression, to avoid inertial and viscoelastic effects. During each cyclic compression protocol, a ccd camera captures the evolution of the beams at a rate of $60~\text{Hz}$ with $\approx 11~\text{pixels/mm}$.

\subsection{Precise definition of phases and transitions}
\begin{figure}[htb]
    \centering
    \includegraphics[]{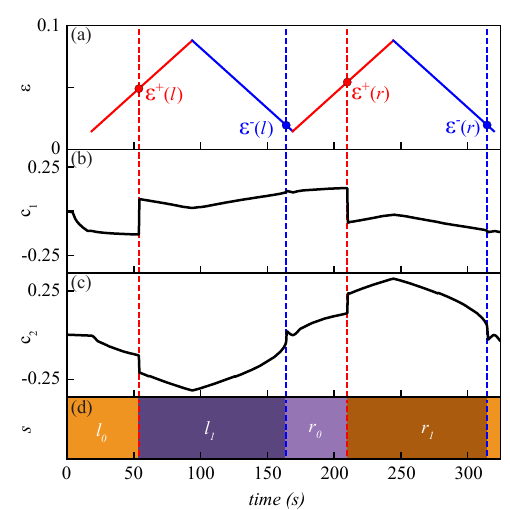}
    \caption{Modal analysis of the beam shape of the toggleron of Fig.~2 main text. 
    (a) Cyclic driving protocol with  irreversible transitions during compression (red) and decompression (blue).
    (b-c) Evolution of the amplitudes of the first and second Euler modes, $c_1,c_2$.
    (d) Corresponding toggerlon phases.
    }\label{fig:modes}
\end{figure}

We now discuss how the toggleron phases,  $s=l_0,~l_1,~r_0$, and $r_1$, can be detected in the real-space configuration of the beam. 
We consider the evolution of the toggleron shown in Fig.~2 of the main text. We extract the dimensional beam shape $X(Z)$ from the images, define rescaled coordinates
$x = X/(L_0\left(1-\varepsilon \right))$ and $z = {Z}/({L_0\left(1-\varepsilon \right)})$, and focus on the  normalized beam shape $x(z)$. We expand the normalized shape using the first six Euler buckling modes \cite{MeulblokEML2025}:
\begin{align}
    x(z) \approx \sum_{i=1}^{6}c_i \xi_i(z)~,
\end{align}
where $\xi_i(z)$ is the $i$-th mode \cite{MeulblokArXiv2025,Pandey14}.
The irreversible transitions are associated to changes of  
the signs of the amplitudes of the first two modes; 
$c_1$ changes sign during the transition at increasing
$\varepsilon$, and $c_2$ changes sign for decreasing
$\varepsilon$ (Fig.~\ref{fig:modes}). Hence, these signs allow to extract the toggleron phase directly from the beam shape (Table.~\ref{tab:phases}). 

\begin{table}[htb]
    \centering
    \begin{tabular}{c|c|c}
            State & $c_1$ & $c_2$ \\ \hline
            $l_0$  & $< 0$ & $\leq 0 $ \\
            $l_1$  & $> 0$ & $> 0 $ \\
            $r_0$  &  $> 0$ & $\geq 0 $ \\
            $r_1$  & $< 0$ & $ < 0 $ \\
    \end{tabular}
    \caption{Relation between the signs of $c_1$ and $c_2$ and toggleron phase. 
    }\label{tab:phases}
\end{table}

\subsection{Design parameters}
We summarize the design parameters of the two coupled togglerons presented in Fig.~4 of the main text. The geometry of the beams is given by $(t_{b1},t_{b2})=(4.0\pm.1~\text{mm}, 3.0\pm.1~\text{mm})$, and the hinge width of the flexible legs of each pusher base is $(w_{L1}, w_{R1}, w_{L2}, w_{R2})=(2.0\pm.1~\text{mm}, 7.0\pm.1~\text{mm}, 2.0\pm.1~\text{mm}, 1.8\pm.1~\text{mm})$  (Fig.~\ref{fig:SI03}). The thickness of the flexible legs $w_p=6.0\pm0.1~\text{mm}$ for all boundaries except for $R2$ which is $w_p=8.0\pm0.1~\text{mm}$.

\begin{figure}[h]
    \centering
    \includegraphics[]{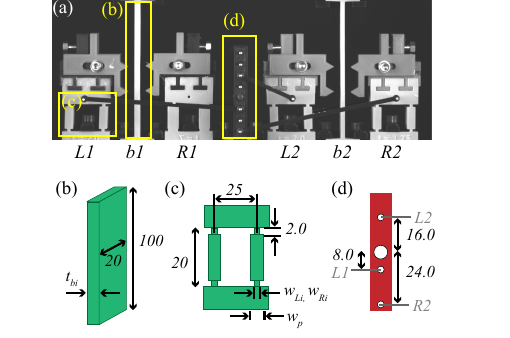}
    \caption{Geometry.
    (a) System indicating the left beam (b), the flexible pusher legs of boundary L1 (c) and central pivot of the coupling mechanism (d).
    (b-d) All dimensions in mm with error $0.1~\text{mm}$; in (d), gray lines denote the connection to the specified pusher.}
    \label{fig:SI0X}
\end{figure}

\section{Movie captions}

\textbf{Movie 1: Toggleron}\\
Evolution of the toggleron shown in Fig.~2 of the main text under cyclic compression. Notice that the movie starts from zero strain and the beam first buckles upon initial compression. After this, six compression cycles are shown, before relaxing back to zero compression. Scale bar is 20~mm.\\

\textbf{Movie 2: $(\tau,T)=(2,2)$ orbit}\\
Evolution of the system of two interacting togglerons for $\varepsilon_M^7$ (see 
Fig.~4e of the main text). After initialization, six compression cycles are shown, before relaxing to zero strain. Scale bar is 20~mm.\\

\textbf{Movie 3: $(\tau,T)=(0,4)$ orbit}\\
Evolution of the system of two interacting togglerons at $\varepsilon_M^6$ (see Fig.~4f of the main text). After initialization, 12 compression cycles are shown, before relaxing to zero strain. Scale bar is 20~mm.\\

\textbf{Movie 4: $(\tau,T)=(1,3)$ orbit}\\
Evolution of the system of two interacting togglerons at $\varepsilon_M^5$ (see Fig.~4g of the main text). After initialization, seven compression cycles are shown, before relaxing to zero strain. Scale bar is 20~mm.\\











\end{document}